\documentstyle[
a4wide]{article}
\begin{document}

\newcommand{\lsim}{\mbox{\raisebox{-.9ex}{~$\stackrel{\mbox{$<$}}{\sim}$~}}}
\newcommand{\gsim}{\mbox{\raisebox{-.9ex}{~$\stackrel{\mbox{$>$}}{\sim}$~}}}

\begin{titlepage}

\title{\Large\bf Trapped Quintessential Inflation}

\author{J.C. Bueno S\'anchez and K. Dimopoulos\\
\mbox{\hspace{1cm}}\\
{\it Physics Department, Lancaster University, Lancaster LA1 4YB, U.K.}\\
\mbox{\hspace{1cm}}}

\date{May, 2006}

\maketitle

\abstract{ \normalsize\noindent 
Quintessential inflation is studied using a string modulus as the inflaton - 
quintessence field. The modulus begins its evolution at the steep part of its 
scalar potential, which is due to non-perturbative effects (e.g. gaugino 
condensation). It is assumed that the modulus crosses an enhanced symmetry 
point (ESP) in field space. Particle production at the ESP temporarily traps 
the modulus resulting in a brief period of inflation. More inflation follows, 
due to the flatness of the potential, since the ESP generates either an 
extremum (maximum or minimum) or a flat inflection point in the scalar 
potential. Eventually, the potential becomes steep again and inflation is 
terminated. After reheating the modulus freezes due to cosmological 
friction at a large value, such that its scalar potential is dominated by 
contributions due to fluxes in the extra dimensions or other effects. The 
modulus remains frozen until the present, when it can become 
quintessence and account for the dark energy necessary to explain the observed 
accelerated expansion.
}

\thispagestyle{empty}

\end{titlepage}

\pagebreak

To everyone's surprise recent observations of high-redshift supernovae have
suggested that the Universe at present is engaging into a phase of accelerated
expansion \cite{SNae}. This result has been confirmed by a number of other 
independent observations such as the distribution of large scale structure 
\cite{lss} and also from combinations of the precise CMB anisotropy data with 
the estimates of the Hubble constant from the Hubble space telescope key 
project \cite{wmap3+hst}. In fact, many 
cosmologists now accept the so-called concordance model of cosmology, according
to which more than 70\% of the Universe content at present is attributed to 
the so-called Dark Energy. Dark Energy is an elusive substance, with pressure 
negative enough to cause the observed acceleration \cite{DE}
(for a recent review see \cite{ed}). The simplest type of Dark 
Energy is a positive cosmological constant $\Lambda$, which however, needs to 
be fine-tuned to an incredibly small value in order to explain the 
observations \cite{L}. 
Hence, despite the fact that considering a non-zero $\Lambda$
is the simplest and by far the easiest to use option, theorists have looked for
alternative explanations, which could explain the observations while setting
\mbox{$\Lambda=0$}, as was originally assumed. A promising idea along this 
direction is to consider that the Universe at present is entering a late-time 
inflationary period \cite{early}. 
The credibility of this option has been enhanced by the 
fact that the generic predictions of the inflationary paradigm in the Early 
Universe are very much in agreement with the observations. The scalar field 
responsible for this late-inflation period is called quintessence because it
is the fifth element after baryons, photons, CDM and neutrinos \cite{Q}. 

Since they are based on the same idea, it was natural to attempt to unify
early Universe inflation with quintessence. Quintessential inflation was thus 
born \cite{quinf,QI,jose,eta}. 
The advantages of this effort are many. The obvious one has to do with
the fact that quintessential inflation models allow the treatment of both
inflation and quintessence within a single theoretical framework, with
hopefully fewer and more tightly constraint parameters. Another practical 
gain is the fact that quintessential inflation dispenses with a tuning problem
of quintessence models; that of the initial conditions for the quintessence 
field. It is true that attractor - tracker quintessence alleviates this 
tuning but the problem never goes away completely. In contrast, in 
quintessential inflation the initial conditions for the late time accelerated 
expansion are fixed in a deterministic manner by the end of inflation. Finally,
a further advantage of unified models for inflation and quintessence is the
economy of avoiding to introduce yet again another unobserved scalar field
to explain the late accelerated expansion.

For quintessential inflation to work one needs a scalar field with a runaway
potential, such that the minimum has not been reached until today and, 
therefore, there is residual potential density, which, if dominant, can cause 
the observed accelerated expansion. String moduli fields appear suitable for 
this role because they are typically characterised by such runaway potentials. 
The problem with such fields, however, is how to stabilise them temporarily, in
order to use them as inflatons in the early Universe. In this letter (see also
Ref.~\cite{ours}) we achieve
this by considering that, during its early evolution our modulus crosses an
enhanced symmetry point (ESP) in field space. As shown in Ref.~\cite{trap},
when this occurs the modulus may be trapped at the ESP for some time. This
can lead to a period of inflation, typically comprised by many sub-periods
of different types of inflation such as trapped, eternal, old, slow-roll, 
fast-roll and so on. After the end of inflation the modulus picks up speed 
again in field space resulting into a period of kinetic density domination,
called kination \cite{kination}. 
Kination is terminated when the thermal bath of the hot big 
bang (HBB) takes over. During the HBB, due to cosmological friction 
\cite{cosmofric}, the modulus
freezes asymptotically at some large value and remains there until the present,
when its potential density becomes dominant and drives the late-time 
accelerated expansion \cite{eta}.

Is is evident that, in order for the scalar field to play the role of 
quintessence, it should not decay after the end of inflation. Reheating, 
therefore should be achieved by other means. One option is gravitational 
particle production as discussed in Ref.~\cite{ford}.
However, because such reheating is typically quite inefficient, in this paper
we assume that the thermal bath of the HBB is due to the decay of some curvaton
field \cite{curv,curv+} as suggested in Refs.~\cite{eta,curvreh}.
Note that the curvaton can be a realistic field, already present in simple
extensions of the standard model (for example it can be a right-handed 
sneutrino \cite{sneu}, a flat direction of the (N)MSSM \cite{mssm,nmssm}
or a pseudo Nambu-Goldstone boson \cite{pngb,orth}
possibly associated with the Peccei-Quinn 
symmetry \cite{PQ} etc.).  Thus, by considering a curvaton we do not
necessarily add an ad~hoc degree of freedom. The importance of the curvaton 
lies also in the fact 
that the energy scale of inflation can be much lower than the grand unified 
scale \cite{liber}.
In fact, in certain curvaton models, the Hubble scale during inflation can be 
as low as the electroweak scale \cite{orth,low}.

String theories contain a number of flat directions which are parametrised by
the so-called moduli fields. Many of such flat directions are lifted by
non-perturbative effects, such as gaugino condensation or D-brane instantons
\cite{Derendinger:1985kk}.
The superpotential, then, is of the form
\begin{equation}
W=W_0+W_{\rm np}\quad \textrm{with} \quad W_{\rm np}=Ae^{-cT}\,,
\end{equation}
where $W_0$ is the tree level contribution from fluxes (which can be 
taken approximately constant), $A$ and $c$ are constants, whose magnitude 
and physical interpretation depends on the origin of the non-perturbative term
(in the case of gaugino condensation $c\lsim 1$), and $T$ is a K\"ahler 
modulus in units of $m_P$, for which
\begin{equation}
  T=\sigma+i\alpha\quad\textrm{with}\quad
  \sigma,\alpha\in {\rm I\!R}.
\end{equation}
Hence, the non-perturbative superpotential $W_{\rm np}$
results in a runaway scalar potential characteristic of string
compactifications. For example, in type IIB compactifications with a single
K\"ahler modulus, $\sigma$ is the so-called volume modulus, which
parametrises the volume of the compactified space. Consequently, in this case,
the runaway behaviour leads to decompactification of the internal manifold.

The tree level K\"ahler potential for a modulus, written in
units of $m_P^2$, is
\begin{equation}
  K=-3\,\textrm{ln}\,(T+\bar{T})\,,
\label{tree}
\end{equation}
and the corresponding supergravity potential is
\begin{equation}
\label{Vnp0}
  V_{\rm np}(\sigma)=\frac{cAe^{-c\sigma}}{2\sigma^2m_P^2}\left[
  \left(1+\frac{c\sigma}{3}\right)Ae^{-c\sigma}+W_0
  \cos(c\alpha)\right]\,.
\end{equation}

To secure the validity of the supergravity approximation we
consider $\sigma>1$. Then, for values of $c\lsim1$, we have
$c\sigma>1$, and we approximate the potential as%
\footnote{The imaginary part $\alpha$ of the $T$ modulus can be
taken such that the ESP lies in a minimum in the direction of
$\alpha$, namely $\cos(c\alpha)=-1$.}
\begin{equation}
\label{Vnp}
 V_{\rm np}(\sigma)\simeq
\frac{cAe^{-c\sigma}}{2\sigma^2m_P^2}
\left(\frac{c\sigma}{3}Ae^{-c\sigma}-W_0\right)\,.
\end{equation}

In order to study the cosmology, we turn to the canonically normalised field 
$\phi$ associated to $\sigma$, which due to Eq.~(\ref{tree}) is given by
\begin{equation}
\label{fs}
  \sigma(\phi)=
\exp\,(\lambda\phi/m_P)\quad\textrm{with}\quad\lambda=\sqrt{2/3}\,.
\end{equation}

Let us assume that the Universe is initially dominated by the above modulus.
The non-perturbative scalar potential for the modulus, shown in 
Eq.~(\ref{Vnp}), 
is very steep (exponential of an exponential), which means that the field
will soon become dominated by its kinetic density. Once this is the case the 
particular form of the scalar potential ceases to be of importance. 

Now, suppose that while rolling towards large values the modulus crosses an
ESP. In string theory compactifications there are distinguished points in
moduli space at which there is enhancement of the gauge symmetries of the 
theory \cite{Hull:1995mz}. This often results in some massive states of the 
theory becoming massless at these points. 
Even though from the classical point of view an ESP
is not a special point, as the modulus approaches it
certain states in the string spectrum become massless \cite{Watson:2004aq}. 
In turn, these massless modes create an
interaction potential that may drive the field back to the
symmetry point. In that way a modulus can become trapped at an ESP
\cite{trap}. The strength of the symmetry
point varies depending on the degree of enhancement of the
symmetry at the particular point; the higher the symmetry the
stronger the force driving the modulus back to the symmetry point.
Such moduli trapping can lead to a period of so-called `trapped inflation'
\cite{trap}, when the trapping is strong enough to make the kinetic density of
the modulus fall below the potential density at the ESP.
However, it turns out that the number of e-foldings of trapped inflation
cannot be very large. Therefore, with respect to cosmology, the main virtue of 
the ESPs relies on their ability to trap the field and
hold it there, at least temporarily, thereby setting the initial conditions 
by confining the modulus to a precise location in field space.
In this sense, quantum effects make the ESP a preferred location.

Let us briefly study the trapping of the modulus at the ESP. Following 
Ref.~\cite{trap}
we assume that around the ESP there is a contribution to the scalar potential
due to the enhanced interaction between the modulus $\phi$ and another field 
$\chi$, which we take to be also a scalar field for convenience. The 
interaction potential is 
\begin{equation}
\label{Vint}
  V_{\rm int}(\phi,\chi)=\frac{1}{2}\,g^2\chi^2\bar\phi^2\,,
\end{equation}
where $\bar{\phi}\equiv\phi-\phi_0$ with $\phi_0$ denoting the value of
the modulus at the ESP and $g$ being a dimensionless coupling constant.
We see that at the ESP the $\chi$ particles become massless. The time
dependence of the effective mass-squared of the
$\chi$ field results in the creation of particles with typical
momentum \cite{trap} 
$k_0\sim(g\dot{\phi}_0)^{1/2}$, where 
$\frac{1}{2}\dot\phi_0^2$ is the kinetic density of the modulus when
crossing the ESP (and the dot denotes derivative with respect to the 
cosmic time $t$). The production takes place when the field is within
the production window $|\phi|<\Delta\phi$, where 
\begin{equation}
\label{window}
\Delta\phi\sim(\dot{\phi}_0/g)^{1/2}\sim k_0/g\,.
\end{equation}

The effective scalar potential, 
$V_{\rm eff}(\phi)=V(\phi)+V_{\rm int}(\phi)$, near the ESP is
\begin{equation}
  V_{\rm eff}(\phi)\approx
  V_0+\frac{1}{2}g^2\langle\chi^2\rangle\bar{\phi}^2
\label{Veff}
\end{equation}
where $V_0\equiv V(\phi_0)$ with $V(\phi)$ being the `background' scalar 
potential. Following Ref.~\cite{trap}
we have $\langle\chi^2\rangle\simeq n_{\chi}/g|\phi|$, 
where $n_{\chi}$ denotes the number density of $\chi$ particles produced 
after the crossing of the ESP. This means that 
$V_{\rm eff}(\phi)\sim V_0+gn_{\chi}|\phi|$ and the field climbs a linear 
potential since $n_{\chi}$ is constant outside the production window. The field
reaches an amplitude $\Phi_1$ given by
\begin{equation}
\label{firstamp}
  \Phi_1/m_P\sim\frac{\dot{\phi}_0^{1/2}}{g^{5/2}m_P}\lsim 1\,,
\end{equation}
which is determined by its initial kinetic density. The upper bound in the 
above is due to the requirement that the modulus does not overshoot the ESP
without being trapped, since for larger values the coupling softens and the 
field, instead of falling back to the
symmetry point, would keep rolling down its potential \cite{Brustein:2002mp}.

After reaching $\Phi_1$ the field reverses direction and crosses the production
window again, generating more $\chi$ particles and, therefore, increasing 
$n_\chi$. Thus, it now has to climb a steeper potential reaching an amplitude
$\Phi_2<\Phi_1$. The process continues until the ever decreasing amplitude 
becomes comparable to the production window shown in Eq.~(\ref{window}). At 
this moment particle production stops. The final value of $n_\chi$ is estimated
as $ n_{\chi}\sim g^{-1/2}\dot{\phi}_0^{3/2}$. After the end of particle 
production, $\langle\chi^2\rangle$ remains roughly constant during an 
oscillation and the modulus continues oscillating in the quadratic interaction 
potential [cf. Eq.~(\ref{Veff})]. Studying this oscillation in 
Ref.~\cite{ours} we found that, due 
to the Universe expansion, the amplitude and frequency decrease as
\begin{eqnarray}
\label{evol}
\Phi(t)\sim\frac{\Delta\phi}{a} & \quad{\rm and}\quad &
\langle\overline{\chi^2}(t)\rangle\sim\left(\frac{k_0}{a}\right)^2=k^2(t)\,,
\end{eqnarray}
where $k(t)$ is the physical momentum and we have normalised the scale factor
$a(t)$ to unity at the end of particle production. When the initial kinetic 
density 
of the modulus is redshifted down to the 
level of $V_0$, trapped inflation begins. 

It is easy to check that the amplitude of the
oscillations at the onset of trapped inflation is
\begin{equation}
\label{trapamp}
\Phi_{\rm i}\sim\frac{\sqrt{m_0m_P}}{g^{1/2}}\,.
\end{equation}
Using this, it can be easily shown that the number of e-foldings of trapped 
inflation, while the modulus oscillates, is \cite{ours}
\begin{equation}
 N_{\rm osc}\sim\ln\left[g^{3/2}\left(\frac{m_P}{m_0}\right)^{1/2}\right]\,,
\end{equation}
where $m_0$ is roughly the Hubble scale during inflation, given by
$m_0\equiv \sqrt{V_0}/m_P\simeq\sqrt 3 H_*$.
Note that $N_{\rm osc}$ is independent of $\dot{\phi}_0$ because the excess of
kinetic energy must be redshifted before trapped inflation can
begin. Therefore, 
once trapped inflation begins the only relevant energy scale is $m_0$.

The redshifting of the frequency of oscillations means that the modulus becomes
lighter. At some point oscillations cease and the field begins to slow-roll
down the interaction potential. The number of slow-roll e-foldings is found to 
be \cite{ours}
\begin{equation}
N_{\rm sr}\sim\frac{4}{3}\ln \frac{1}{g}\,.
\end{equation}
The field value during this phase of slow-roll trapped inflation is roughly 
$|\bar\phi|\sim\Phi_{\rm min}$, where 
\begin{equation}
\label{minamp}
\Phi_{\rm min}\sim m_0/g^2\,
\end{equation}
is the minimum amplitude of the oscillations.

Finally, because the interaction potential becomes
increasingly depleted, the phase of slow-roll of the trapped modulus comes to
an end when the field becomes dominated by its quantum fluctuations. 
After this moment the field drives a period of eternal inflation, which is 
oblivious to the scalar potential. During this phase the field value performs
a one dimensional random walk in field space, of step determined by the 
Hawking temperature $H_*/2\pi$; corresponding to an e-folding of inflation.
Hence, after $N$ e-foldings we have 
$\sqrt{\langle\phi^2\rangle}\sim\frac{H_*}{2\pi}\sqrt{N}$. 
The field value obeys
\begin{equation}
|\bar\phi(N)|\sim m_0/g^2+m_0\sqrt{N}\,.
\end{equation}

The above results are true under the assumption that trapped inflation 
continues uninterrupted. However, since the interaction potential becomes 
gradually depleted, there will be a moment when 
$V'_{\rm int}(\phi)\lsim |V'(\phi)|$, in which the dynamics of the modulus 
begins to be determined by the `background' scalar potential $V(\phi)$ rather
than the interaction potential (where the prime denotes derivative with 
respect to $\phi$). 
Thus, the stages of trapped inflation can be
summarised as follows \cite{ours}:

\begin{itemize}
\item
{\bf Trapping:} After crossing
the ESP, the field starts to oscillate around it if 
{$|V^{\prime}(\Phi_1)|<gn_{\chi}$}, where 
$n_{\chi}\sim g^{3/2}\dot{\phi}_0^{3/2}$ is the number
density of $\chi$ particles after the first crossing.

\item
{\bf Trapped inflation:} 
Trapped inflation occurs whenever 
{$|V^{\prime}(\Phi_{\rm i})|\lsim g^{5/2}(m_0m_P)^{3/2}$}.

\item
{\bf Slow-roll:} 
The oscillations of the
field give way to slow-roll motion if 
{$|V^{\prime}(\Phi_{\rm min})|\lsim m_0^3/g^2$}.

\item
{\bf Eternal inflation:} 
A phase of eternal inflation occurs when
{$|V^{\prime}(\Phi_{\rm min})|\lsim g^2m_0^3$}.
This phase lasts until the field is sufficiently far away from the
ESP so that the scalar potential $V(\phi)$ becomes the driving force again.
\end{itemize}

Because ESPs are fixed points of the 
symmetries, by definition, the scalar potential is flat at these points. Hence,
we have
\begin{equation}\label{eq:firstderiv}
V^{\prime}_0\equiv V^{\prime}(\phi_0)=0\,.
\end{equation}
The above means that the ESP is located either at a local extremum (maximum or 
minimum) or at a flat inflection point of the scalar potential, where
\mbox{$V_0'=V_0''=0$} with \mbox{$V_0''\equiv V''(\phi_0)$}. 
This means that the presence of an ESP is expected to deform the 
non-perturbative
scalar potential. To model the ESP we, therefore, introduce a phenomenological 
contribution $V_{\rm ph}$ to the non-perturbative scalar potential such 
that $V(\phi)=V_{\rm np}(\phi)+V_{\rm ph}(\phi)$ (see Fig.~1). We require that 
$V_{\rm ph}$ becomes negligible far away from the ESP so that the 
steepness of $V_{\rm np}$ is recovered. To be able to perform analytic 
calculations we have chosen the following form for $V_{\rm ph}$:

\vspace{-1cm}

\input{epsf}

\begin{center}

\begin{figure}
\begin{picture}(10,10)
\put(
-4,-190%
){
\leavevmode
\hbox{%
\epsfxsize=%
5.8in
\epsffile{
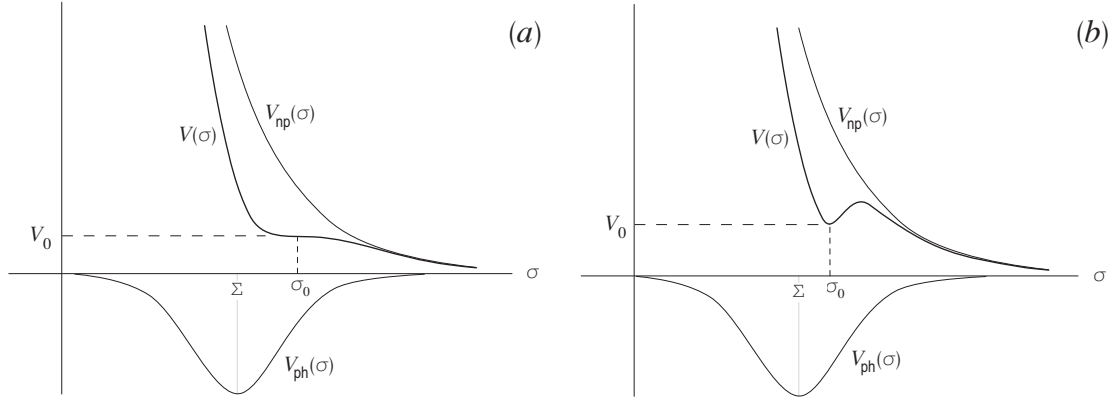
}}
}

\end{picture}

\vspace{7cm}

\caption{\footnotesize
The deformation 
of $V_{\rm np}(\sigma)$ at the ESP as modelled through the use of
our choice for $V_{\rm ph}(\sigma)$. Case~(a) corresponds to a flat inflection 
point, while Case~(b) corresponds to a local minimum.
}
\end{figure}

\end{center}
\begin{equation}
\label{Vph} 
 V_{\rm ph}(\sigma)=-\frac{M}{\textrm{cosh}^q(\sigma-\Sigma)}\,,
\end{equation}
where $M$ is some density scale and $\Sigma, q$ are positive constants.
It can be shown that the functional form of $V_{\rm ph}$ is not important
\cite{ours}. The results, in fact, depend only on the following parameters
\begin{equation}
m_\phi^2\equiv V''(\phi_0)\quad{\rm and}\quad
V^{\textrm{\tiny(3)}}_0\equiv V'''(\phi_0)\,.
\end{equation}
Combining Eqs.~(\ref{Vnp}) and (\ref{Vph}) one can show that 
$V^{\textrm{\tiny(3)}}_0$ can be parametrised as follows
\begin{equation}
\label{xi0}
|V^{\textrm{\tiny{(3)}}}_0|\approx
2(\lambda\mu)^3(c\sigma_0)^3\xi^2\frac{m_0^2}{m_P}\,,
\end{equation}
where $\sigma_0\equiv\sigma(\phi_0)$,
$\mu=2,1$ if Eq.~(\ref{Vnp}) is dominated either by the first
or second term, respectively. The parameter $\xi$, defined by
\begin{equation}
\xi\equiv\frac{\sqrt{q}}{c\mu}>1\,,
\label{xi}
\end{equation}
accounts for the strength of the symmetry point: the smaller the
value of $\xi$ is, the larger the inflationary plateau generated
by the ESP becomes, and therefore, the stronger the deformation 
due to the ESP is. The lower bound on $\xi$ is due to the requirement that
$V_{\rm ph}$ (i.e. the deforming effect of the ESP) becomes negligible far 
away from the ESP.

Before moving on let us discuss briefly the particular case where our 
Kh\"ahler 
modulus is the volume modulus of type IIB compactifications. In this case the 
ESP must be located within the non-perturbative region of string theory. In 
this region, the volume of the compact space (measured in string units) is 
${\cal V}_{\rm st}\ll1$. In terms of the volume modulus $T$ and the string 
coupling 
$g_s$ we have ${\cal V}_{\rm st}\sim g_s^{3/2}(\textrm{Re}\,T)^{3/2}$, where
we take $\textrm{Re}\,T\equiv\sigma$. Therefore, to locate the ESP within
the non-perturbative region we require $\sigma_0< g_s^{-1}$. In this case
$V^{\textrm{\tiny{(3)}}}_0$ may be parametrised as \cite{ours}
\begin{equation}\label{eq:paramv3}
 V^{\textrm{\tiny(3)}}_0=-2(2\lambda)^3\zeta^2\frac{m_0^2}{m_P}\,,
\end{equation}
where $\zeta$ is defined as
\begin{equation}
\label{zeta}
  \zeta\equiv\frac{\sigma_0\sqrt{q}}{2}>1\,.
\end{equation}
The physical meaning of $\zeta$ is identical to that of $\xi$ in 
Eq.~(\ref{xi}).

Let us study first the case where the ESP creates a flat inflection point
such that $V'_0=V''_0=0$. In this case the inflationary dynamics,
arising from the scalar potential $V(\phi)$ in the neighbourhood
of the ESP, can be accounted for by approximating the scalar potential
as
\begin{equation}
V(\phi)\approx
V_0+ \frac{1}{3!}V^{\textrm{\tiny{(3)}}}_0\bar{\phi}^3.
\end{equation}
To quantify our parameter space we consider that
$m_0\sim 1$~TeV. For a generic Kh\"ahler
modulus we take $(c\sigma_0)\sim 10$, $c\lsim 1$ and
$g\simeq0.1$. Then the following cases are possible \cite{ours}:

\begin{itemize}
\item If $g^2m_0<|V^{\textrm{\tiny(3)}}_0|$ then
the field is unable to drive a long-lasting period of inflation because 
it escapes trapping before the end of the oscillations.

\item If $g^6m_0\lsim|V^{\textrm{\tiny(3)}}_0|<g^2m_0$ then
the slow-roll after the end of oscillations is continued with
another phase of slow-roll over the scalar potential $V(\phi)$.
The number of e-foldings attainable is
$N_{{\rm sr}_1}\sim g^2m_0/|V^{\textrm{\tiny (3)}}_0|\lsim g^{-4}$.
For our choice of parameter values we find that this case corresponds to
$10^5<\xi^2<10^9$. Then the maximum number of e-foldings in this case is 
$N_{{\rm sr}_1}\lsim 10^4$.

\item If $|V^{\textrm{\tiny(3)}}_0|<g^6m_0$ then
the slow-roll phase after the end of oscillations is
continued with a phase of eternal inflation, which, typically, lasts for
$N_{\rm et}\gsim m_0/|V_0^{\textrm{\tiny (3)}}|>g^{-6}$ e-foldings.
The end of eternal inflation is followed by another phase of
slow-roll over the scalar potential $V(\phi)$ with number of e-foldings
$N_{{\rm sr}_2}\sim\sqrt{m_0/|V^{\textrm{\tiny (3)}}_0|}\gsim g^{-3}$.
For our choice of parameter values we find that this case corresponds to
$1<\xi^2<10^5$. The maximum number of eternal inflation e-foldings is
$N_{\rm et}\sim 10^{10-11}$. Similarly, the maximum number of
e-foldings corresponding to the second slow-roll period is 
$N_{{\rm sr}_2}\sim 10^5$.
\end{itemize}
In order not to overshoot the ESP, the interaction coupling has to satisfy the 
bound $g\gsim 10^{-3}$. In the interval $10^{-3}\lsim g\lsim10^{-2}$ the 
resulting slow-roll phase needs not to be preceded by eternal inflation, while 
in the interval $10^{-2}<g\lsim1$ eternal inflation has to occur.

Another important bound on the parameters stems from the requirement that the
density perturbations due to the inflaton modulus are not excessive compared
to the observations. This requirement translates into the bound \cite{ours}
\begin{equation}
\xi^2\lsim
\frac{10^{-4}}{N^2(c\sigma_0)^3}\,\frac{m_P}{m_0}\sim10^4\,,
\end{equation}
where we take $N\sim10^2$. 
Fortunately, eternal inflation is followed by enough slow-roll inflation to 
encompass the cosmological scales. 
In the case where $\sigma$ is the volume 
modulus one can show that, the requirement
that the inflaton's perturbations are not excessive translates into 
\cite{ours}
\begin{equation}
\zeta^2\lsim\frac{10^{-4}}{N_{{\rm sr}_1}^2}\frac{m_P}{m_0}\,,
\end{equation}
which is quite easily satisfied.

We move on now to considering the case when the ESP results in an extremum in
the scalar potential. This time we have $V''_0=m_\phi^2\neq 0$ and the scalar
potential can be approximated as
\begin{equation}
V(\phi)\simeq V_0+\frac{1}{2}m_{\phi}^2\bar{\phi}^2-
\frac{1}{2}\frac{m_{\phi}^2}{\bar{\phi}_{\rm t}}\bar{\phi}^3\,,
\end{equation}
where $\bar{\phi}_{\rm t}$ is defined by $V(\phi_0)=V(\phi_{\rm t})$,
i.e. it is the value of the modulus where the scalar potential has the
same height as at the ESP. It is evident that in the case where the ESP 
corresponds to a local minimum \{maximum\} $\bar{\phi}_{\rm t}$ is positive
\{negative\}. The value of $\bar{\phi}_{\rm t}$ depends on the magnitude of
$V^{\textrm{\tiny(3)}}_0$. Using Eq.~(\ref{xi0}) we obtain
\begin{equation}
\label{ft}
\bar{\phi}_{\rm t}\sim\frac{\eta\,\xi^{-2}}{(c\sigma_0)^3}\,m_P\,,
\end{equation}
where $\eta=(m_\phi/m_0)^2$ is the second slow-roll parameter at the ESP.

Suppose at first that the ESP lies at a local minimum. In this case after 
trapping the field has to tunnel out, as in old inflation. 
We assume that the field emerges at $\approx\phi_{\rm t}$ after
tunnelling. Then, the number of e-foldings of slow-roll after bubble 
nucleation is $N_{\rm sr}\sim(m_0/m_{\phi})^2\gsim N_{\rm H}$, where 
$N_H\gsim 60$ is the number of e-foldings necessary to produce a flat
Universe at scales of at least $10^2H^{-1}$ so that the flatness and horizon 
problems be solved. Therefore, the mass of the field at the ESP must be
$m_{\phi}\lsim\frac{1}{8}\,m_0$. Note that, only if there is enough slow-roll
inflation after bubble nucleation can we hope to overcome the graceful exit 
problem of old inflation.

Another bound is due to the requirement that the modulus does not generate
excessive density perturbations. This results in the bound \cite{ours}
\begin{equation}
\xi^2\lsim\frac{10^{-4}}{N_{\rm sr}^2(c\sigma_0)^3}
\frac{m_P}{m_0}\sim10^4\,,
\end{equation}
with a maximum number of e-foldings $N_{\rm sr}\sim10^3$. 

Now suppose that the ESP lies at a local maximum. If the tachyonic mass of the 
field  is $|m_{\phi}|\lsim m_0$ the field may drive a period of inflation with
total number of e-foldings given by \cite{ours}
\begin{equation}
N_{\rm tot}\sim\frac{1}{F_{\phi}}
\ln\left(\frac{|\eta|g^2}{(c\sigma_0)^3\xi^2}\frac{m_P}{m_0}\right)\,,
\end{equation}
where $F_\phi\equiv\frac{3}{2}\left(\sqrt{1+\frac{4}{3}|\eta|}-1\right)$. When 
$|m_{\phi}|\ll m_0$ inflation is slow-roll and \mbox{$F_\phi\approx|\eta|$}. 
When $|m_{\phi}|\sim m_0$ inflation is fast-roll and 
\mbox{$F_\phi={\cal O}(1)$}.
The requirement that the inflaton modulus does not produce excessive density 
perturbations results in the bound \cite{ours}
\begin{equation}
\xi^2\lsim 
\frac{|\eta|^2\exp(-N_{\rm H}F_{\phi})}{10^4(c\sigma_0)^3}\frac{m_P}{m_0}\,.
\end{equation}
For the above to be consistent with the parameter space for $\xi$ we need
$10^{-4}\lsim |\eta|\lsim0.25$, which is a reasonable range.

After the end of inflation the field rolls away of the ESP. Soon the influence 
of the ESP on the scalar potential diminishes and 
$V(\phi)\approx V_{\rm np}(\phi)$ (i.e. $V_{\rm ph}$ becomes negligible).
The steepness of $V_{\rm np}$ results in the kinetic domination of the modulus
density. As a result a period of kination takes place, during which the field
equation is $\ddot\phi+3H\dot\phi\simeq 0$, which suggests that the density
of the Universe scales as $\rho\simeq\frac{1}{2}\dot\phi^2\propto a^{-6}$ 
\cite{kination}. During kination the scalar field is oblivious 
of the particular form of the scalar potential as long as the latter does not
inhibit kination. Kination is terminated when the density of 
an oscillating curvaton field, or of radiation due to the curvaton's decay
overtake the kinetic density of the modulus \cite{curvreh}. 
For the purposes of this work both
of these possibilities are roughly equivalent. Therefore, we assume that the
curvaton decays before it dominates the Universe. In this case the end of
kination is also the onset of the HBB, i.e. it corresponds to the reheating of 
the Universe, with reheating temperature 
$T_{\rm reh}\sim g_*^{-1/4}\sqrt{H_{\rm reh}m_P}$, with $g_*\sim 10^2$ being 
the number of relativistic degrees of freedom and $H_{\rm reh}$ being the 
Hubble parameter at reheating.\footnote{In quintessential inflation, typically,
reheating occurs late, which requires the curvaton to decay in advance, in 
order to account for baryogenesis.}

As shown in Ref.~\cite{eta} (see also \cite{cosmofric})
after the onset of the HBB the rolling scalar field is subject to cosmological
friction which asymptotically freezes the field at a value $\phi_F$ 
corresponding to 
\begin{equation}
\label{sF}
\sigma_F\sim\left(\frac{m_0\,m_P}{g_*T_{\rm reh}^2}\right)^{2/3}.
\end{equation}
Because $\sigma_F>1$ the scalar potential is no longer dominated by 
$V_{\rm np}$ shown in Eq.~(\ref{Vnp}). Instead it is dominated by a 
contribution of the form
\begin{equation}
\label{exps}
V(\sigma)\simeq\frac{C_n}{\sigma^n}
\quad\Rightarrow\quad 
V(\phi)\simeq C_ne^{-b\phi/m_P}\,,
\end{equation}
with $b=n\lambda=\sqrt{\frac{2}{3}}\,\,n$. The modulus remains frozen at the
value shown in Eq.~(\ref{sF}) until the present. This guarantees that there is
no dangerous variation of fundamental constants despite the fact that the 
modulus is not stabilised at a local minimum of the scalar potential. 

At present, if the modulus is to account for the required dark energy, its 
potential density has to satisfy the coincidence requirement 
$V(\sigma_F)\simeq\Omega_\Lambda\rho_0$, where $\Omega_\Lambda\simeq 0.73$ is
the density parameter of dark energy and $\rho_0$ is the critical density at 
present. Using Eqs.~(\ref{sF}) and (\ref{exps}) one finds that the coincidence
requirement demands
\begin{equation}
\label{Treh}
T_{\rm reh}\sim\sqrt{m_0\,m_P}
\left(\frac{\rho_0}{C_n}\right)^{\sqrt{3/8n^2}}.
\end{equation}
Because reheating has to occur before nucleosynthesis takes place, the above 
results in the bound
\begin{equation}
\label{Cbound}
C_n\lsim\rho_0
\left(\frac{\sqrt{m_0\,m_P}}{T_{\sc bbn}}\right)^{2n\sqrt{2/3}},
\end{equation}
where $T_{\sc bbn}\sim 1$~MeV is the temperature at nucleosynthesis. This
is an important constraint on the scalar potential in Eq.~(\ref{exps}).

Another important constraint is due to the possibility of excessive 
gravitational wave generation. The spectrum of gravitational waves in models 
of quintessential inflation features a spike at high frequencies due to the
stiff equation of state during kination \cite{eta,grav}.
In order for these gravitons not to disturb nucleosynthesis one has to impose 
the bound \cite{ours}
\begin{equation}
m_0\lsim (4\times 10^2)^{3/8}(m_PT_{\sc bbn})^{1/2}\sim 10^5\,\textrm{TeV}\,.
\end{equation}
Thus, we see that inflation has to take place at lower energies than the grand
unified scale. This supports the use of a curvaton field for the generation of
the density perturbations because low-scale inflation is easier to attain 
\cite{orth,liber,low}.

The scalar potential considered in Eq.~(\ref{exps}) 
may have a multitude of origins.
For example, if we use the volume modulus, we may consider a number of 
$\overline{D3}-$branes located at the tip
of a Klebanov-Strassler throat.
The contribution $\delta V$, proportional to the warp factor, is
 \cite{Giddings:2001yu}
\begin{equation}
  \delta V(\sigma)\sim e^{-8\pi K/3Mg_s}\frac{m_P^4}{\sigma^2}\equiv
  C_2/\sigma^2\,,
\end{equation}
where $M$ and $K$ quantify the units of RR and NS three-form fluxes.
To satisfy Eq.~(\ref{Cbound}) we must have $C_2^{1/4}\lsim10^{-20}m_P$.
This can be obtained with
a choice of fluxes $K/Mg_s\gsim22$. Taking $g_s=0.1$, only
approximately twice as many units of $K$ flux as those of $M$ flux
are needed.

It is also possible to consider fluxes of gauge fields on
$D7-$branes \cite{Burgess:2003ic}. 
In this case, the scalar potential obtains a contribution
\begin{equation}
\delta V\sim \frac{2\pi E^2}{\sigma^3}\equiv C_3/\sigma^3\,,
\end{equation}
where $E$ depends on the strength of the gauge fields considered.
The constraint in Eq.~(\ref{Cbound}) requires now
$C_3^{1/4}\lsim 10^{-15}m_P\sim 1$~TeV.

Another kind of corrections introducing an uplifting term in the
scalar potential $V(\phi)$, changing its structure at large volume
and breaking the no-scale structure, are $\alpha^{\prime}$ corrections
\cite{Becker:2002nn}.
The uplifting term is due to a corrected Kh\"ahler potential
\cite{Westphal:2005yz}
\begin{equation}
K=K_0-2\textrm{ln}
\left(1+\frac{\hat{\xi}}{2(2\textrm{Re}\,T)^{3/2}}\right)\,,
\end{equation}
where $K_0$ is the tree-level Kh\"ahler
potential shown in Eq.~(\ref{tree}), and $\hat{\xi}=-\frac{1}{2}\zeta(3)\chi
e^{-3\varphi/2}$ where $\chi$ is the E\"uler number of the
internal manifold, and $e^{\varphi}=g_s$ is the string coupling.
In this case the uplifting term $\delta V$ is
\begin{equation}
\delta V\sim\frac{\hat{\xi}\,W_0^2}{\sigma^{9/2}}\equiv
\frac{C_{9/2}}{\sigma^{9/2}}\,.
\end{equation}
Thus, the bound in Eq.~(\ref{Cbound}) with $m_0\sim1$ TeV requires
$C_{9/2}^{1/4}\lsim 10^{-7}m_P$, i.e. it corresponds to the intermediate scale
$C_{9/2}^{1/4}\lsim10^{11}$ GeV.

In closed string theory, in addition to the usual translational
modes, vibrational modes of closed strings wrapping around the
compact manifold are also present. The risk is that the presence
of this modes may disturb nucleosynthesis.
The mass of these excitations follows an inverse relation to the
radius of the compactification \cite{Conlon:2005ki}
$m_{\rm KK}\sim m_s/R\sim g_s \,m_P/{\cal V}^{2/3}_{\rm st}$,
where $R$ is the radius of the compactified space defined by the
volume of the Calabi-Yau manifold ${\cal V}_{\rm st}\sim R^{6}$.
These modes decay at the temperature $T_{\rm KK}\sim(g_s/\sigma_F)^{3/2}m_P$.
Demanding $T_{\rm KK}>T_{\sc bbn}$ constrains $\sigma_F$. Using Eq.~(\ref{sF}) 
we obtain a lower bound on the reheating temperature
\begin{equation}
T_{\rm reh}>(g_*g_s^3)^{-1/4}\sqrt{T_{\sc bbn}\,m_0}\,.
\end{equation}
For $m_0\sim1$~TeV, this bound results in $T_{\rm reh}>1$ GeV. 
Combining the above with Eq.~(\ref{Treh}) we get
 a stronger bound on $C_{9/2}$, which reads
$C_{9/2}^{1/4}\lsim10^{-12}m_P\sim10^6$GeV.

The future of the modulus after unfreezing depends on the steepness of the 
potential, or equivalently the value of $b$ in Eq.~(\ref{exps}). 
\begin{itemize}
\item
For $b\leq\sqrt 2$, the modulus dominates the Universe for ever, leading to 
eternal acceleration. This results in future horizons, which pose a problem
for the formulation of the S-matrix in string theory \cite{S}.

\item
For $\sqrt 2< b\leq\sqrt 3$, the modulus dominates the Universe but results 
only in a brief accelerated expansion period. Such is the fate of the 
$n=2$ case. 

\item
For $\sqrt 3<b\leq\sqrt 6$ the modulus does not dominate the Universe, albeit
causing a brief period of accelerated expansion. After this period the modulus
density remains at a constant ratio with the background matter density.
This is the fate of the $n=3$ case.

\item
For $\sqrt 6<b\leq 2\sqrt 6$, the modulus also does not dominate the Universe, 
but still causes a brief period of accelerated expansion. Afterwards the
modulus rolls fast down the quintessential tail of the scalar potential with
its density approaching asymptotically kinetic domination (and subsequently
freezing at a larger value than $\sigma_F$). This is the fate of the $n=9/2$
case.

\item
For $b>2\sqrt 6$ the modulus after unfreezing does not cause any accelerated 
expansion and so cannot be used as quintessence. As in the previous case the
modulus rolls fast down the quintessential tail of the scalar potential with
its density approaching asymptotically kinetic domination (and subsequently
freezing at a larger value than $\sigma_F$). This case corresponds to $n>6$.
\end{itemize}
The brief acceleration period caused by the unfreezing modulus is due to the
fact that the modulus oscillates around an attractor solution \cite{oscil}.
This solution does not result to acceleration, but the oscillations of the 
system around it have been found to cause brief periods of accelerated 
expansion, especially just before unfreezing \cite{jose}. 
As shown in Ref.~\cite{cline}
brief acceleration occurs if $\sqrt 2<b\leq 2\sqrt 6$, which corresponds to
the range $\sqrt 3<n\leq 6$. In all cases the modulus eventually approaches
$+\infty$ which corresponds to a supersymmetric ground state. If $\sigma$
is the volume modulus then this final state leads to decompactification of 
the extra dimensions. 

In summary, we have studied quintessential inflation using a string modulus as 
the inflaton - quintessence field. In our model the periods of accelerated
expansion both in the early Universe and at present are due to the dynamical 
evolution of the compactified extra dimensions, which, in the intermediate
epoch, remain frozen due to cosmological friction.
Our inflaton modulus is initially kinetic density
dominated as it rolls down its steep non-perturbative potential, due to
to gaugino condensation or D-brane instantons. We assumed that the modulus
crosses an enhanced symmetry point (ESP), which results in substantial particle
production. The produced particles generate a contribution to the scalar 
potential which can trap the modulus at the vicinity of the ESP, giving rise to
a limited period of trapped inflation. The latter can be followed by a longer
period of inflation after the modulus is released from the trapping. This 
period is due to the fact that the scalar potential is deformed around the ESP,
so that the ESP is located at either an extremum of a flat inflection point. 
We have studied carefully all possible cases and showed that a variety of
inflationary phases may take place, such as eternal, old, slow-roll and 
fast-roll inflation. We then investigated the parameter space which produces
enough inflation to solve the horizon and flatness problems and found that
this is indeed possible for reasonable values of the model parameters, mostly
determined by the extend of the effect of the ESP on the form of the scalar 
potential. After the end of inflation the modulus rolls away from the ESP and 
it becomes kinetically dominated again. Thus, inflation is followed by a period
of kination which is interrupted by the domination of the decay products of a 
curvaton field. A curvaton has been employed not only to generate the required 
density perturbations but also because its decay products are able to reheat 
the Universe. After reheating the evolution of the modulus is dominated by
cosmological friction due to the thermal bath of the hot big bang. 
Consequently, the modulus approaches asymptotically a fixed value, where it 
becomes stabilised (frozen) despite the fact that it does not lie in a local
minimum of the potential. The freezing value of the field is now such that 
the scalar potential for the canonicaly normalised modulus is dominated by 
an exponential contribution, which can be 
due to a number of string induced effects, such as fluxes  or D-branes in the 
extra dimensions or $\alpha'$-corrections to the K\"ahler potential.
The modulus remains frozen until the present, when its potential density 
becomes comparable to the density of the Universe once more. At this moment 
the modulus is expected to unfreeze and play the role of quintessence. We have
shown that the exponential potential can satisfy the coincidence requirement
for reasonable values of the parameters. In this model quintessence is about to
unfreeze at the present time. Therefore, in the past our model does not offer
predictions different from $\Lambda$CDM. However, it may conceivably be linked 
to the intriguing possibility that some of the fundamental constants, such as 
the fine-structure constant, may begin to vary at present. This issue 
lies beyond the scope of this work but deserves future consideration.

\bigskip

KD would like to thank Steve Giddings for stimulating discussions and the
Aspen Center for Physics for the hospitality.

\end{document}